\documentclass{aa}  

\usepackage{graphicx}

\usepackage{txfonts}
\usepackage[colorlinks=true,citecolor=cyan]{hyperref}
\begin{document}

   \title{Hierarchical accretion flow from the G351 infrared dark filament to its central cores}


   \author{Beuther H.
          \inst{1}
          \and
          Olguin F.A.,
          \inst{2,3,4}
          \and
          Sanhueza P.,
          \inst{5,6} 
          Cunningham N.
          \inst{7}
          \and
          Ginsburg A.
          \inst{8}
          }

 \institute{\inst{1} Max Planck Institute for Astronomy, Königstuhl 17, 69117 Heidelberg, Germany, email: name@mpia.de\\
   \inst{2} Center for Gravitational Physics, Yukawa Institute for Theoretical Physics, Kyoto University, Kitashirakawa Oiwakecho, Sakyo-ku, Kyoto 606-8502, Japan\\
   \inst{3} National Astronomical Observatory of Japan, National Institutes of Natural Sciences, 2-21-1 Osawa, Mitaka, Tokyo 181-8588, Japan\\
   \inst{4} Institute of Astronomy, National Tsing Hua University, Hsinchu 30013, Taiwan\\
   \inst{5} Department of Earth and Planetary Sciences, Institute of Science Tokyo, Meguro, Tokyo, 152-8551, Japan\\
   \inst{6} National Astronomical Observatory of Japan, National Institutes of Natural Sciences, 2-21-1 Osawa, Mitaka, Tokyo 181-8588, Japan\\
   \inst{7} SKA Observatory, Jodrell Bank, Lower Withington, Macclesfield SK11 9FT, United Kingdom\\
   \inst{8} Department of Astronomy, University of Florida, PO Box 112055, Gainesville, USA
             }


 
  \abstract
   {Quantifying the accretion flow from large cloud scales down to individual protostars is a central ingredient to the understanding of (high-mass) star formation.}
   {We characterize and quantify this multi-scale flow for a prototypical high-mass star-forming region.}
   {In a multi-scale analysis from parsec to $\sim$50\,au scales, we combined multiple single-dish and interferometric observations to study the gas flow from large-scale sizes of several parsec (Mopra) via intermediate-scale filamentary gas flows (ALMA-IMF) to the central cores (ALMA DIHCA and configuration 10 data). The highest-resolution multi-configuration ALMA dataset achieved a spatial resolution of $0.027''\times 0.022''$ or 50\,au.}
   {This multi-scale study allows us to follow the gas from the  environment of the high-mass star-forming region ($\sim$2\,pc) via intermediate-scale ($\sim$0.25\,pc) filamentary gas flows down to the innermost cores within the central few 1000\,au. The intermediate-scale filaments connect spatially and kinematically to the larger-scale cloud as well as the innermost cores. We estimate a filamentary mass inflow rate around $10^{-3}$\,M$_{\odot}$\,yr$^{-1}$, feeding into the central region that hosts at least a dozen mm cores. While the flow from the cloud via the filaments down to 10$^4$\,au appears relatively ordered, within the central 10$^4$\,au the kinematic structures become much more complicated and disordered. We speculate that this is caused by the interplay of the converging infalling gas with feedback processes from the forming central protostars.}
   {This multi-scale study characterises and quantifies the hierarchical gas flow from clouds down to the central protostars for a prototypical infrared dark cloud with several embedded cores at an unprecedented detail. While comparatively ordered gas flows are found over a broad range of scales, the innermost area exhibits more disordered structures, likely caused by the combination of inflow, outflow and cluster dynamical processes.}


   \maketitle

\section{Introduction}
\label{intro}

The formation of stars requires gas flowing from the parental large molecular cloud down to the central protostars. A quantification on how the gas flows over a broad range of spatial scales remains one of the major problems in observational star formation research. Cloud dynamics and mass flows have been studied separately toward molecular clouds (e.g., \citealt{henshaw2013,peretto2013,barnes2018,beuther2020}), more filamentary cluster-forming structures (e.g., \citealt{kirk2013,peretto2014,henshaw2014,liu2015,beuther2015b,chen2019,barnes2023,sandoval-garrido2024}), or more recently also the flows toward individual cores within high-mass star-forming regions (e.g., \citealt{sanhueza2021,wells2024}) and even disks (e.g., \citealt{maud2017,olguin2023}). In low-mass star-forming regions, recently the discussion focused a lot on gas streamers that may feed individual cores and disks (e.g., \citealt{pineda2023,valdivia2023,gieser2024,kueffmeier2024}). Following slightly different approaches, statistical studies of larger samples of (high-mass) star-forming regions searching for asymmetric line profiles revealed evidence for global collapse in such environments (e.g., \citealt{wu2003,fuller2005,jackson2019}). However, so far investigations are lacking for individual regions how the gas flows from the parsec-scale clouds toward the inner 50\,au around the protostar. 

Here, we are aiming at exactly that for the high-mass star-forming filament G351 that hosts at the very center a hot molecular core G351.77-0.54. To study the kinematics from the large-scale infrared dark filament, we combine single-dish data from the MALT90 survey \citep{jackson2013} with interferometric mosaic observations from the Atacama Large Millimeter/submillimeter Array (ALMA) covering the immediate environment of the central cluster (ALMA-IMF, \citealt{motte2022,ginsburg2022,cunningham2023,sandoval-garrido2024}) to very-high-spatial-resolution ALMA data toward the central cores \citep{beuther2017b,beuther2019,olguin2021,taniguchi2023}.

\begin{figure*}[ht]
	\includegraphics[width=0.99\linewidth ,keepaspectratio]{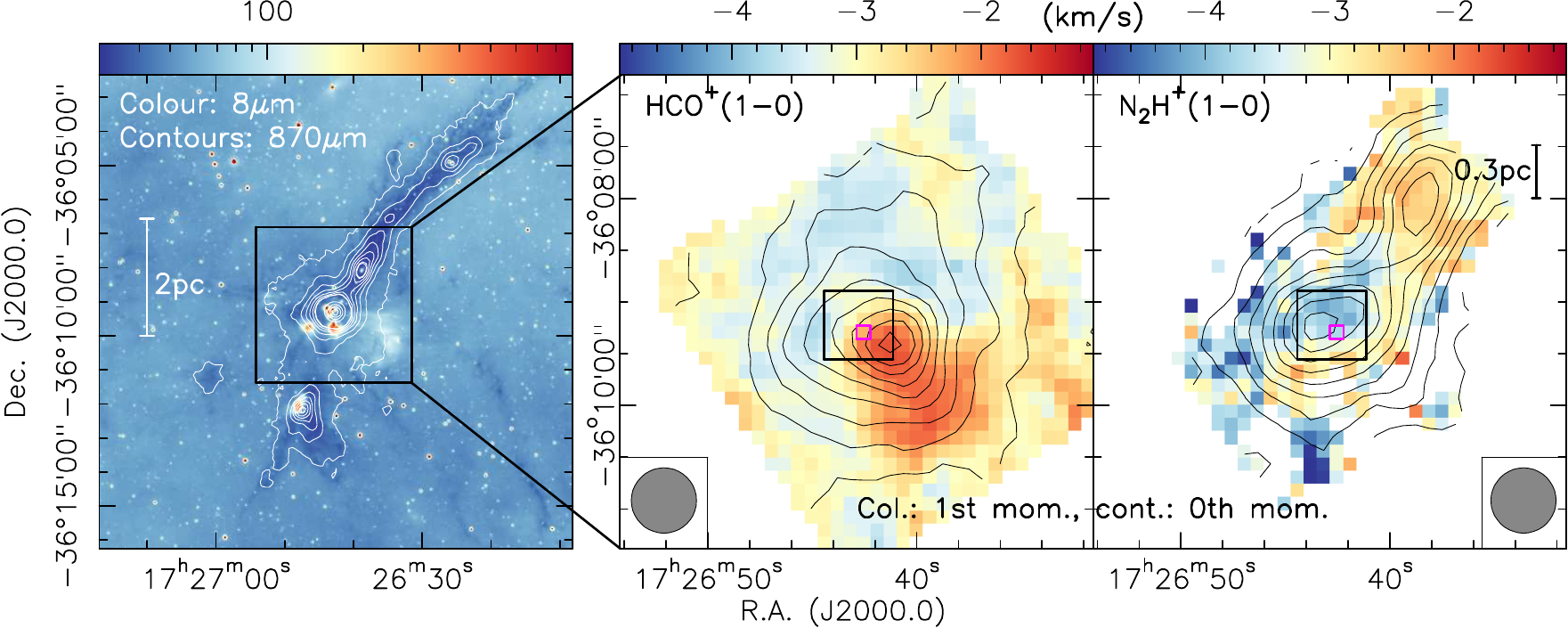}
	\caption{Overview of the G351 region. The left panel shows in color-scale the Spitzer 8\,$\mu$m emission \citep{churchwell2009} and in contours the ATLASGAL 870\,$\mu$m emission \citep{schuller2009}. Contour levels start at the 4$\sigma$ level of 240\,mJy\,beam$^{-1}$. The middle and right panels present the molecular emission from the HCO$^+$(1--0)  and N$_2$H$^+$(1--0) lines observed with the MALT90 survey \citep{jackson2013}. In both cases, the color-scale shows the 1st moment maps (intensity-weighted peak velocities), and the contours present the integrated line emission or 0th moment from 5 to 95\% of the respective peak emission. For N$_2$H$^+$(1--0), the 1st moment map is from the isolated hyperfine component shifted by +8.0\,km\,s$^{-1}$ to the $v_{\rm{lsr}}$. Since the rms increases toward the map edges, we masked the emission outside the 5\% 0th moment map for the N$_2$H$^+$(1--0) map.  The black and magenta boxes in the middle and right panels outline the areas shown in the following ALMA-IMF and ALMA central core images. Linear scale-bars are shown in the left and right panels, and the MALT90 beams are presented in the bottom corners.}
	\label{g351_large}
\end{figure*}

The target of our investigation is the well-know high-mass star-forming region G351.77-0.54 (also known as IRAS\,17233-3606) that is part of the larger-scale filamentary infrared dark cloud IRDC G351.77-0.53 (left panel of Fig.~\ref{g351_large}, e.g., \citealt{leurini2011b}). The region is very line rich and hosts class II CH$_3$OH masers (e.g., \citealt{norris1993,walsh1998,leurini2011,beuther2017b,beuther2019,cunningham2023,bonfand2024}). While the region drives strong outflows (e.g., \citealt{leurini2009,leurini2013,klaassen2015,towner2024}), rotational motions have been identified in highly excited NH$_3$ and CH$_3$CN emission \citep{beuther2009c,beuther2017b}. The absolute distance to the region has been debated, and values typically between 1.0 and 2.2\,kpc have been reported (e.g., \citealt{norris1993,leurini2011,motte2022}). A recent analysis of Gaia parallaxes of likely members of the region find a distance of $\sim$2.0$\pm 0.14$\,kpc \citep{reyes2024}. We use that distance in the following. Scaling the luminosity of the region estimated by \citet{urquhart2018} to 2\,kpc distance, G351.77-0.54 has a luminosity of $\sim$8.2$\times10^4$\,L$_{\odot}$. Similarly, if we scale the mass derived by \citet{urquhart2018} to this new distance, the main G351.77-0.54 star-forming region has a mass of $\sim$2065\,M$_{\odot}$ over an area with radius $\sim$0.55\,pc. For comparison, analysis of highly excited CH$_3$CN emission around the main central mm peak at $0.06''$ resolution shows that the data are consistent with a 10\,M$_{\odot}$ central protostar \citep{beuther2017b}. One should keep in mind that this region is forming a cluster and already within the central $\sim$10$^4$\,au at least 12 mm cores have been resolved \citep{beuther2019}.

\section{Data and observations}

\subsection{Large-scale and single-dish data}

For the largest spatial scales on the order of the parsec-long parental filament, we resorted to the mid-infrared and and submm Galactic plane surveys Spitzer-GLIMPSE \citep{churchwell2009} and ATLASGAL \citep{schuller2009}. We used the Spitzer 8\,$\mu$m image with a spatial resolution of $\sim$2$''$. The corresponding ATLASGAL 870\,$\mu$m data have a beam size of $19''$.

Since we are interested in the kinematics, on scales of several parsec around the high-mass star-forming region G351 we used the Millimetre Astronomy Legacy Team 90\,GHz (MALT90) survey that mapped with the Mopra 22\,m telescope a $3'\times3'$ area around the central G351 region in several spectral lines \citep{foster2011,foster2013,jackson2013}. We mainly used the datacubes in the HCO$^+(1-0)$ (89.188526\,GHz) and N$_2$H$^+(1-0)$ (93.173772\,GHz) lines with a spatial and spectral resolution of $38''$ and 0.11\,km\,s$^{-1}$. The typical rms noise was 0.25\,K ($T_A^*$) in a 0.11\,km\,s$^{-1}$ channel. The original MALT90 N$_2$H$^+(1-0)$ map was shifted by roughly $23''$ to the east compared to the ATLASGAL 870\,$\mu$m map. However, morphologically both maps agree well with the filamentary structure extending toward the northwest. Therefore, we shifted all MALT90 maps by 3 pixels or $22.9''$ to the west to account for this difference, most likely caused by pointing of the single-dish telescope.

\subsection{Atacama Large Millimeter/submm Array (ALMA) data}

\subsubsection{ALMA-IMF mosaic}

The ALMA-IMF program is an ALMA cycle 5 large program that observed 15 high-mass protoclusters covering a span of evolutionary stages. A program overview and data reduction details are presented in \citet{motte2022} and \citet{ginsburg2022}. The regions were typically observed as mosaics in band 6 (1.3\,mm) and band 3 (3\,mm). For G351.77, the corresponding mosaic sizes were $132''\times 132''$ and $190''\times 180''$, respectively. From the many observed spectral lines, we focus only on two, in particular on the DCN(3--2) and H$_2$CO(3--2) lines. The spectral resolution for the two lines was 0.34 and 0.17\,km\,s$^{-1}$, respectively. The 1.3\,mm data were observed in one ALMA 12\,m array configuration (C43-3) and the 7\,m array to also cover larger spatial scales. For more details about the data reduction and imaging procedures, we refer to \citet{ginsburg2022}. The spatial resolution and 1.3\,mm continuum $1\sigma$ rms values are $1.1''\times 0.8''$ and 0.6\,mJy\,beam$^{-1}$, respectively. The spectral line datacubes were recently released by \citet{cunningham2023}. The $1\sigma$ rms values for the DCN(3--2) and H$_2$CO(3--2) lines were 14.4 and 15.8\,mJy\,beam$^{-1}$ at 0.34 and 0.17\,km\,s$^{-1}$ resolution, respectively \citep{cunningham2023}. 

\begin{figure*}[ht]
	\includegraphics[width=0.99\linewidth ,keepaspectratio]{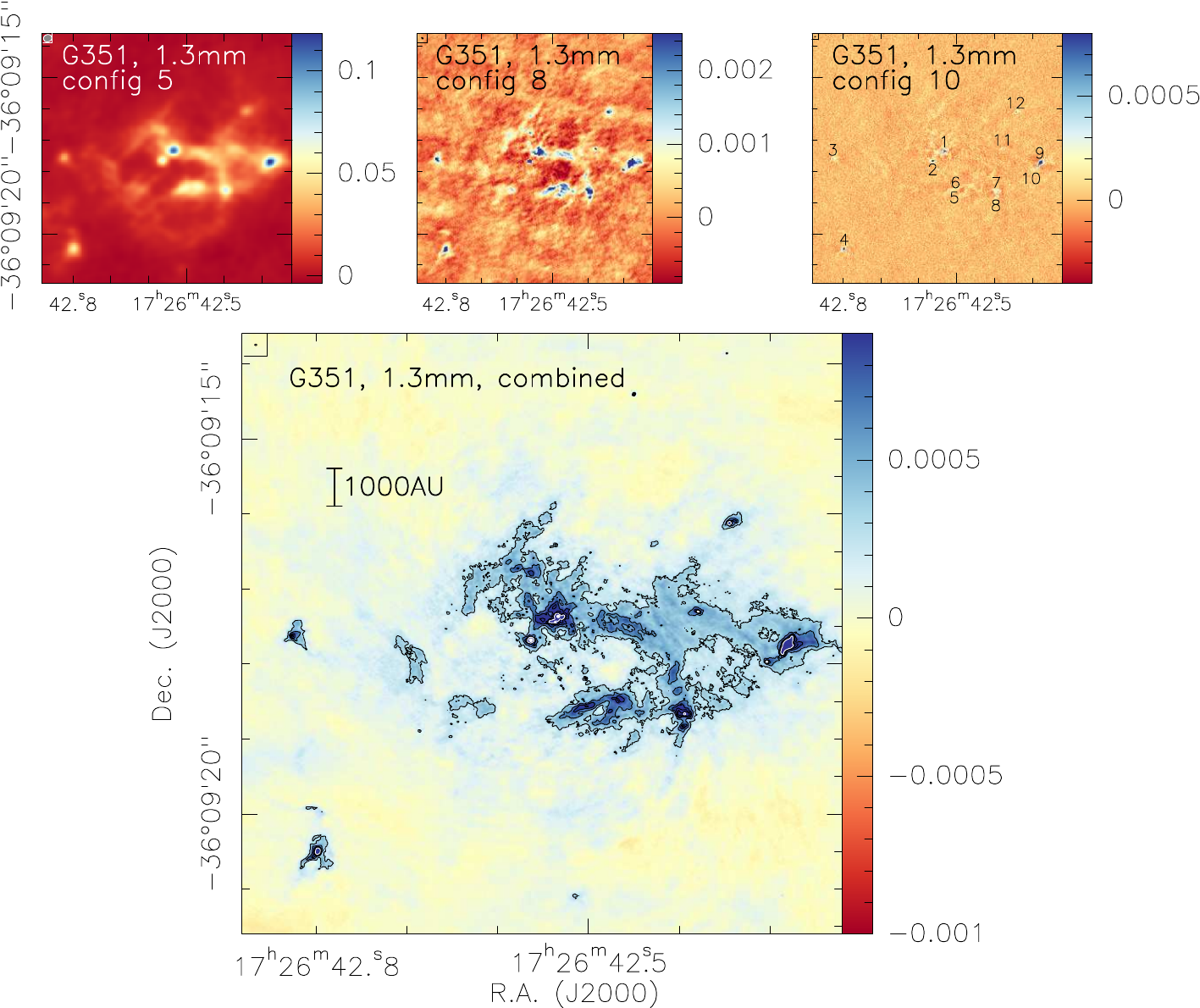}
	\caption{Compilation of the different ALMA 1.3\,mm continuum data for the central G351 hot core region (little magenta box in Fig.~\ref{g351_large}). The top-three panels present from left to right the configuration 5 and 8 data from the DIHCA survey \citep{olguin2021,olguin2022,taniguchi2023} and the highest resolution configuration 10 data \citep{beuther2019}, all in R.A.\,(J2000) and Dec.\,(J2000). In this panel, also the mm cores identified in \citet{beuther2019} are labeled. The bottom panel then presents the combined merged data showing all spatial scales present in each individual dataset. The contour levels go from the 5 to 30$\sigma$ level in steps of 5$\sigma$ and then continue 80$\sigma$ steps (1$\sigma$=60\,$\mu$Jy\,beam$^{-1}$). The corresponding synthesized beams are shown in the top-left of each panel. A linear scale-bar is shown in the bottom panel.}
	\label{continuum}
\end{figure*}

\subsubsection{ALMA zoom toward the center}

For the highest spatial resolution zoom into the central region of G351, we combined the data from two projects conducted with ALMA in the 1.3\,mm band. The intermediate-spatial resolution observations in ALMA configurations 5 and 8 were conducted as part of the DIHCA project (Digging into the interior of hot cores with ALMA) led by PI P.~Sanhueza (project ID: 2016.1.01036.S). Early results from DIHCA can be found in, for example, \citet{olguin2021,olguin2022}, \citet{taniguchi2023}, \citet{li2024}, and \citet{ishihara2024}. The corresponding highest-angular-resolution ALMA configuration 10 data, also in the 1.3\,mm band, were observed in project ID 2015.1.00496.S and were published in \citet{beuther2019}. While these very-high-resolution data (25\,mas$\times$20\,mas) resolve the small-scale structure of the region well, they clearly resolve out all extended emission \citep{beuther2019}. 

To overcome this extreme spatial filtering issue, we now combined the DIHCA configuration 5 and 8 data with the separately observed configuration 10 data. For details about the configuration 10 data calibration, self-calibration and imaging, we refer to \citet{beuther2019}. The corresponding DIHCA configuration 5 and 8 data were observed with four spectral windows, two in the lower sideband (216.9--218.7\,GHz and 219.0--221.0\,GHz) and two in the upper sideband (231.0--233.0\,GHz and 231.0--233.0\,GHz). Calibration and imaging was conducted in CASA. Self-calibration was done separately for the configuration 5 and 8 data with four iterations of phase self-calibration for each configuration. The first iteration used the configuration 10 data as model for a better spatial alignment of the three datasets. Then phase-only self-calibration was done with three decreasing solution intervals: for the configuration 8 data the length of the scan (solint='inf'), 30 and 15\,sec, for the configuration 5 data 30, 20 and 10sec. The signal-to-noise ratio between the first and last self-calibration iterations improved from 145 to 200 and from 30 to 108 for the configuration 8 and 5 data, respectively. To evaluate the scales traced by each configuration, we first imaged all three datasets separately, and then conducted a joint deconvolution of all configurations together with a Briggs robust weighting of 0. Figure \ref{continuum} presents an overview of the individually and the jointly imaged datasets. Scientific details are discussed in section \ref{small-scale}. The angular resolution of the individual configuration continuum datasets are $0.27''\times 0.23''$ (position angle PA=$-76^o$), $0.057''\times 0.034''$ (PA=$48^o$) and $0.021''\times 0.015''$ (PA=$-74^o$), respectively. The angular resolution of the jointly deconvolved continuum image in the bottom-right panel of Fig.~\ref{continuum} is then $0.027''\times 0.022''$ (PA=$88^o$). This image now combines the high spatial resolution with the recovery of also the larger spatial scales (see also Section \ref{small-scale}). The $1\sigma$ rms of the continuum and spectral line data (we mainly focus on H$_2$CO$(3_{2,2}-2_{2,1})$ at 218.476\,GHz, also the combination of all three configurations) is 60\,$\mu$Jy\,beam$^{-1}$ and 0.7\,mJy\,beam$^{-1}$ (in 1.0\,km\,s$^{-1}$ channels). The synthesized beam of the H$_2$CO$(3_{2,2}-2_{2,1})$ line with a Briggs robust weighting of 0 is $0.035''\times 0.03''$.

\section{Results}

\begin{figure*}[ht]
	\includegraphics[width=0.99\linewidth ,keepaspectratio]{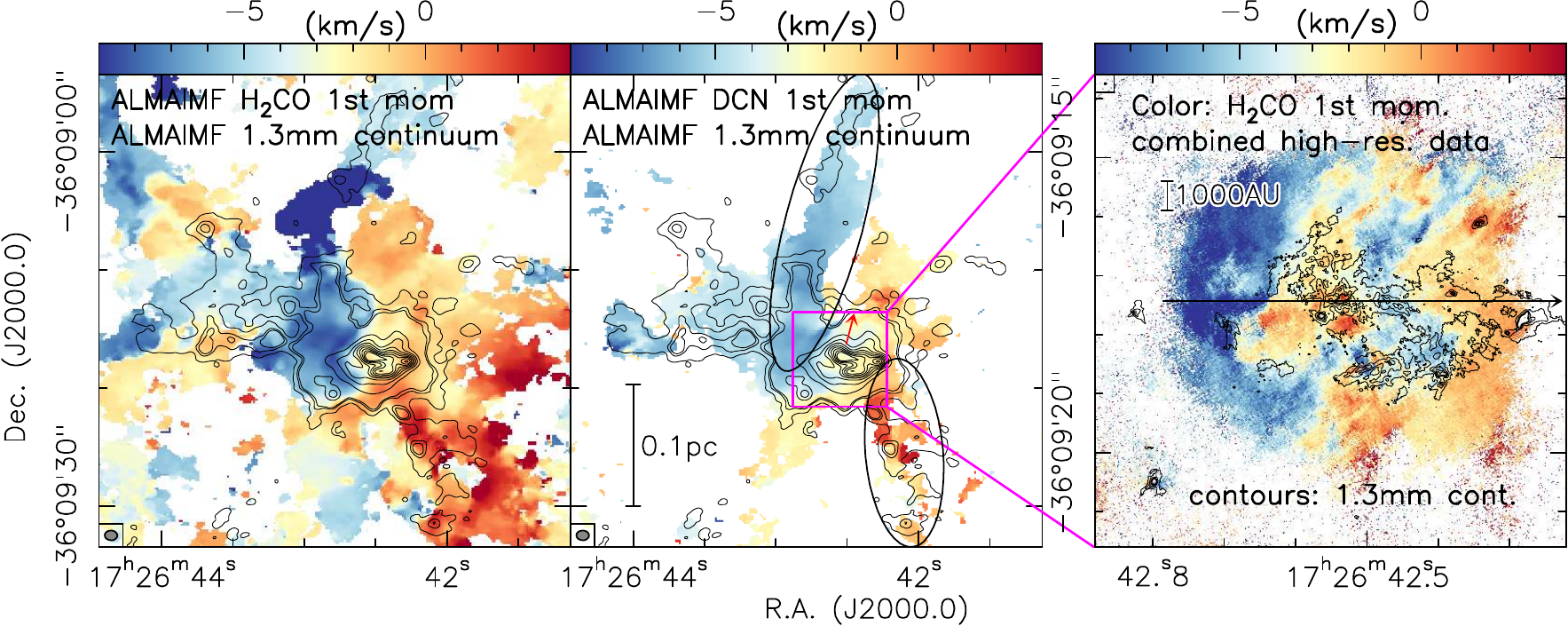}
	\caption{Zoom-in plots from the more central region in G351. The two left panels show the ALMA-IMF data \citep{motte2022,ginsburg2022,cunningham2023} from the first zoom-region (black box in Fig.~\ref{g351_large}). The color-scale in the left and middle panels are the 1st moment maps in H$_2$CO and DCN. The contours show the 1.3\,mm continuum emission starting at 4$\sigma$ and continue in 8$\sigma$ steps (1$\sigma$=0.6\,mJy\,beam$^{-1}$). The ellipses in the middle panel outline the areas used for the mass flow rate estimates, and the red arrow shows the direction of the red-shifted outflow lobe from \citet{beuther2017b}. The right panel then shows the final zoom-in (magenta boxes in middle panel and Fig.~\ref{g351_large}) data from the DIHCA and high-resolution studies \citep{olguin2021,beuther2019} with the H$_2$CO 1st moment map in color-scale and the 1.3\,mm continuum data in contours (5 to 25$\sigma$ in 5$\sigma$ steps with 1$\sigma$=60\,$\mu$Jy\,beam$^{-1}$). The arrow in the right panel marks the direction of the position-velocity cut in Fig.~\ref{pv_h2co}. Linear scale-bars are shown in the middle and right panels, synthesized beams are presented in the bottom- and top-left corners of the respective panels.}
	\label{ALMA-IMF}
\end{figure*}

\subsection{Large-scale structure and dynamics}
\label{largescale}

Figure \ref{g351_large} (left panel) shows how the central G351 high-mass star-forming region is located in the middle of a filamentary infrared dark cloud (IRDC), typically considered as sites of early star formation (e.g., \citealt{egan1998,pillai2006,peretto2009,beuther2020}). While the IRDC largely presents a pristine environment, the 8\,$\mu$m image already shows localized ongoing star formation toward the central region that was discussed in Sect.~\ref{intro}. If one focuses more on the central region and the corresponding kinematics, the middle and right panels present the integrated emission and the 1st moment maps (intensity-weighted peak velocities) of the HCO$^+(1-0)$ and N$_2$H$^+(1-0)$ lines. While the integrated N$_2$H$^+$ emission recovers well the filamentary north-eastern structure also seen in the 870\,$\mu$m emission (Fig.~\ref{g351_large} left panel), the HCO$^+$ emission is less structured and more spherically shaped around the central region. These structural slightly different appearances may well be optical depth effects where HCO$^+$ has a higher optical depth and traces more of the diffuse environmental cloud emission whereas N$_2$H$^+$ as an optically thin high-density tracer follows better the dust continuum emission (e.g., \citealt{sanhueza2012}). Independent of that, both spectral line data reveal a velocity gradient roughly in north-south direction across the region. 
We will get back to that structure in the higher-resolution data discussed below.

\subsection{Intermediate-scale filamentary structure and kinematics}
\label{intermediate-scale}

Zooming into the region more, the left and middle panels of Fig.~\ref{ALMA-IMF} present the kinematics and spatial structure of the gas and dust at roughly 2000\,au resolution for the H$_2$CO(3--2) and DCN(3--2) lines. While the central part appears elongated roughly in the east-west direction, one clearly identifies filamentary structures leading toward the central region from the northern and southern direction. More precisely, the blue-shifted gas connects to the central area at its north-eastern edge, whereas the red-shifted gas connects at the south-western side. Qualitatively speaking, these blue- and red-shifted filaments are the high-spatial-resolution counterparts to the large-scale blue-red velocity gradient across the entire star-forming region discussed in Sect.~\ref{largescale} (see also Fig.~\ref{g351_large}). While one can identify potentially even more gas feeding structures, for instance, the blue-shifted extension to the east (Fig.~\ref{ALMA-IMF} left and middle panel), in the following we concentrate on the most pronounced filamentary structures to the north and south marked by the ellipses in the middle panel of Fig.~\ref{ALMA-IMF}.

Although the general inclination of the large- as well as small-scale filamentary structure is not known, the data are consistent with filamentary gas streams feeding the central $\sim$10$^4$\,au of this star-forming region. In such a filamentary streaming picture, one can quantify mass flow rates $\dot{M}$ following, for example, \citet{kirk2013}, \citet{henshaw2014}, and \citet{wells2024}:

$$\dot{M} = \frac{M_{\rm fil} \Delta v}{l_{\rm fil}} \times \frac{1}{{\rm tan}(i)}$$

with $M_{\rm fil}$ and $l_{\rm fil}$ the mass and length of the filament, and $\Delta v$ the velocity difference from one end of the filament to the central core region. The masses $M_{\rm fil}$ can be estimated from the 1.3\,mm continuum data assuming optically thin dust emission following \citet{hildebrand1983} or \citet{schuller2009}. The factor $\frac{1}{{\rm tan}(i)}$ reflects the unknown inclination angle $i$ of the filamentary structures. \citet{wells2024} investigated the influence of this inclination uncertainty on the estimated flow rates, and they find a spread in the distribution of flow rates of roughly 1 order of magnitude at full-width-half maximum.

For the mass estimates, we assume a dust opacity $\kappa = 0.9$\,cm$^2$g$^{-1}$ (\citealt{ossenkopf1994}, MRN dust at $10^6$\,cm$^{-3}$ and 1.3\,mm), a gas-to-dust mass ratio of 150 \citep{draine2003} and a temperature of 30\,K \citep{urquhart2018}. The velocity differences $\Delta v$ are measured from the DCN 1st moment map between the northern and southern ends of the filaments (marked by the ellipses in Figure \ref{ALMA-IMF} middle panel) with respect to the central velocity of $\sim -3.6$\,km\,s$^{-1}$ ($v_{\rm LSR}$ from \citealt{leurini2011}). The filament lengths $l_{\rm fil}$ are also measured along the main long axis of the ellipses in Fig.~\ref{ALMA-IMF}. Table \ref{flowrates} lists the parameters used for our estimates. While all individual parameters are associated with their own uncertainties, the dominating systematic uncertainty appears to be the unknown inclination angle (see above). Hence, the flow rates are associated with an uncertainty of roughly 1 order of magnitude.

\begin{table}[ht]
\caption{Masses and flow rates}
\label{flowrates}
\begin{center}
\begin{tabular}{l|rrrr}
\hline
Filament & $M_{\rm fil}$ & $\Delta v$ & $l_{\rm fil}$ & $\dot{M}$\\
         & (M$_{\odot}$) & (km\,s$^{-1}$) & (1000\,au) & (10$^{-3}$M$_{\odot}$\,yr$^{-1}$)\\
\hline
North  & 86 & 2.2 & 52 & 0.77\\
South  & 39 & 3.3 & 32 & 0.86\\
\hline
\end{tabular}
\end{center}
\end{table}

Using these parameters, we find mass flow rates close to 10$^{-3}$\,M$_{\odot}$\,yr$^{-1}$ over scales of $\sim$0.25\,pc for both filamentary structures. These filaments are feeding a central region of roughly 10$^4$\,au size that harbors at least 12 cores (Fig.~\ref{continuum}). Hence, these flow rates should be considered as cluster-feeding gas flow rates. A recent N$_2$H$^+$ study of the region also found inflow signatures with inflow rates around $\sim$10$^{-3}$ \,M$_{\odot}$\,yr$^{-1}$ \citep{sandoval-garrido2024}, similar to our findings here. For comparison, absorption line measurements toward one of the central mm peak positions revealed central infall rates onto that specific source mm2 between $10^{-4}$ and $10^{-3}$\,M$_{\odot}$\,yr$^{-1}$ \citep{beuther2017b,beuther2019}. 

If protostellar growth continues on comparably long (free-fall) timescales, the gas feeding the cores must be replenished at the rate it falls into those cores.  For the G351 region, we find that the accretion flow onto the central region is high enough to re-fill the reservoir feeding the central accreting protostars.
We can estimate the free-fall time-scale of the region from the single-dish ATLASGAL data. Using the mass and size of the region (Sect.~\ref{intro}), following \citet{stahler2005}, the free-fall time-scale of the region is $\sim1.6\times 10^5$\,yrs. A steady accretion flow of $10^{-3}$\,M$_{\odot}$\,yr$^{-1}$ over that time allows to channel roughly 160\,M$_{\odot}$ toward the central regions which is reasonable for forming high-mass stars within a few hundred thousand years (e.g., \citealt{mckee2003}).

Comparing our results with infall and accretion flow studies of high-mass star-forming regions, other investigations typically found global infall rates or flow rates along filamentary structures in the regime of $10^{-4}$ to $10^{-3}$\,M$_{\odot}$\,yr$^{-1}$ (e.g., \citealt{fuller2005,peretto2013,henshaw2014,wyrowski2016,chen2019,sandoval-garrido2024,wells2024}), which is consistent with our finding for G351. These high-mass star-forming region flow rates are also considerably higher than streamer flow rates found around low-mass regions (typically around $10^{-6}$\,M$_{\odot}$\,yr$^{-1}$, e.g., \citealt{pineda2020,valdivia2022}).

One may also ask whether this parsec-scale accretion flow may be driven by local gravity or whether large-scale convergent gas flows may contribute? To really study large cloud-scale converging gas flows, one would need to investigate even larger-scale data of less dense and also atomic gas which is beyond the scope of this paper. However, on sub-parsec scales as discussed here and shown in Fig.~\ref{ALMA-IMF}, the red- and blue-shifted filamentary structures north and south of the central cores clearly converge toward these central structures. In that sense, one may call these converging filamentary gas streams. Nevertheless, our data do not allow us to differentiate whether these converging gas streams are caused soley by the gravitational potential of the region or whether externally driven converging gas flows may contribute to the picture.

\begin{figure*}[ht]
	\includegraphics[width=0.99\linewidth ,keepaspectratio]{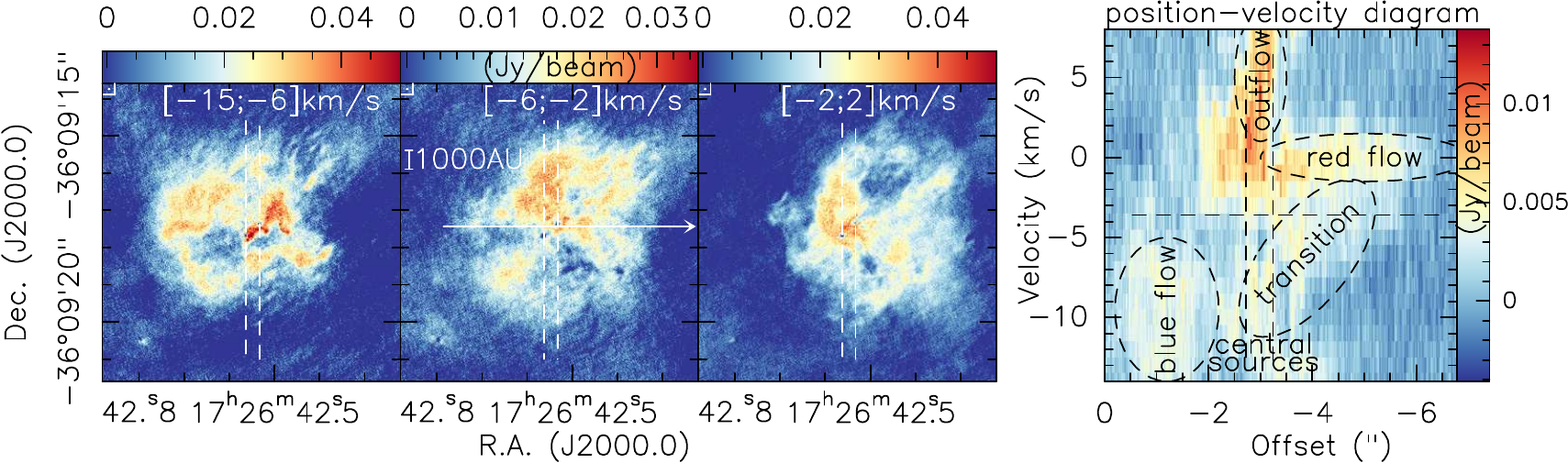}
    \caption{Velocity structure of the central region. The left three panels show integrated H$_2$CO$(3_{2,2}-2_{2,1})$ maps for different velocity regimes as marked. The right panel presents a corresponding position-velocity cut from east to west along the line shown in the right panel of Fig.~\ref{ALMA-IMF} and in the 2nd panel from left here. The dashed vertical lines in all panels mark the R.A.~positions of the two central cores mm1 and mm2 (right and left dashed lines, respectively), where mm1 also drives a molecular outflow in northwest-southeast direction \citep{beuther2019}. The horizontal dashed line in the pv-diagram shows the $v_{\rm LSR}\sim -3.6$\,km\,s$^{-1}$ from \citet{leurini2011}. Other features are marked and labeled. }
	\label{pv_h2co}
\end{figure*}

\subsection{Small-scale kinematics around the central cores}
\label{small-scale}

Moving to the smallest scales and the inner $\sim$10$^4$\,au of this high-mass star-forming region, the high-spatial-resolution study by \citet{beuther2019} identified 12 mm-continuum emission peaks within this central area (top-right panel of Fig.~\ref{continuum}). If one compares that high-resolution image to the lower-resolution configuration 5 image (top-left panel in Fig.~\ref{continuum}), all these sources can be re-identified, just now embedded in a larger-scale, almost filamentary environment. The configuration 8 image (top-middle panel of Fig.~\ref{continuum}) shows as expected a mixture of the two extremes. Finally, the jointly deconvolved combined image of all three configurations reveals the best of the "three worlds" (bottom-big panel of Fig.~\ref{continuum}). At almost the same very high spatial resolution ($0.027''\times 0.022''$) as the configuration 10-only image, one now finds all the small-scale structures embedded in the filamentary environment of the parental gas clump. These data are an excellent example of the power of combining multi-configuration interferometry data into final images and datacubes.

Since we are interested here mainly in the kinematics of the region, the right panel of Fig.~\ref{ALMA-IMF} shows the velocity structure of the region via a 1st moment map of the H$_2$CO$(3_{2,2}-2_{2,1})$ line which is one of the lines the DIHCA and high-resolution data have in common. This line has an upper energy level $E_u/k$ of 68\,K with a critical density of $\sim 3\times 10^{6}$\,cm$^{-3}$. Hence, it is a good tracer of the gas dynamics in the inner center of this star-formation complex.

Inspecting this 1st moment map, one clearly sees that blue- and red-shifted emission are found at the eastern and western side of the central region, consistent with the kinematics of the intermediate-scale filaments discussed previously (Sect.~\ref{intermediate-scale} and Fig.~\ref{ALMA-IMF}). However, in addition to the velocity structure at the eastern and western edges, one does not identify a smooth velocity gradient across but one gets the impression of almost a velocity oscillation across the region varying regularly between blue- and redshifted gas (Fig.~\ref{ALMA-IMF} right panel). 

To have a closer look at this velocity structure, Fig.~\ref{pv_h2co} presents H$_2$CO$(3_{2,2}-2_{2,1})$ integrated intensity maps over selected velocity regimes, and a position-velocity (pv) diagram in the east-west direction across this central region (marked by an arrow in Fig.~\ref{ALMA-IMF} right panel and Fig.~\ref{pv_h2co} the 2nd from left panel). The three velocity regimes cover the important blue- and red-shifted parts as well as velocities in a transition regime between the blue- and red-shifted flow. One clearly sees that significant fractions of the blue- and red-shifted gas are located in the east and west of the region, respectively (left and 3rd panel from left in Fig.~\ref{pv_h2co}), corresponding to the inflow zones discussed in the previous section from the intermediate-scale inflowing gas (Fig.~\ref{ALMA-IMF}). The velocities closer to the $v_{\rm LSR}\sim -3.6$\,km\,s$^{-1}$ show less preferred orientation (2nd panel in Fig.~\ref{pv_h2co}).  Nevertheless, the blue- and red-shifted emission is not found exclusively in the east and west but also more distributed over the region. 

A different way to visualize that is the position-velocity diagram conducted along a cut from east to west (right panel in Fig.~\ref{pv_h2co}). The pv-diagram clearly confirms the complex velocity structure, also showing that moment analysis can partly be misleading in complex kinematic environments. While the 1st moment map in Fig.~\ref{ALMA-IMF} exhibits structure mainly between $-9$ and plus a few km\,s$^{-1}$, the position-velocity diagram reveals a much broader velocity spread from $<-15$\,km\,s$^{-1}$ to $>+8$km\,s$^{-1}$. Consistent with the 1st moment and integrated velocity maps, the position-velocity diagram again shows that the gas at the eastern side is largely blue-shifted and at the western side largely red-shifted. Around the two central cores mm1 and mm2 (marked by the dashed lines in Fig.~\ref{pv_h2co}), all velocities are found. This may partly also be due to the molecular outflow driven by mm1 in northwest-southeast direction (Fig.~\ref{ALMA-IMF} middle panel, \citealt{beuther2017b,beuther2019}). Regarding the velocity oscillation indicated by the 1st moment map, that is less well identified in the position velocity diagram (nor in the three integrated velocity maps). While the transition from blue-shifted gas at the eastern edge to red-shifted gas close to the central mm cores mm1 and mm2 is also evident in the position-velocity diagram, moving further west there is always red-shifted gas around 0\,km\,s$^{-1}$. There is just an additional more blue-shifted gas component between $-5$ and $-14$\,km\,s$^{-1}$ at offsets $-3.5''$ to $-4.0''$ in the pv-diagram that gets weaker again going further to the west (labeled "transition" in Fig.~\ref{pv_h2co} right panel).

Therefore, the gas structure within the central $\sim$10$^4$\,au contains blue- and red-shifted gas at the eastern and western edges, most likely associated with the intermediate-scale gas stream. However, going closer to the center, we cannot identify clear velocity structures or gradients leading to one or the other core. It rather appears that the central area contains several different gas structures that interact. This interacting dynamics can be caused by several processes, in particular the infalling filamentary and converging gas streams, and feedback processes of the forming protostars, specifically from the main central molecular outflow around mm1. In addition to this, other cores in the region may also drive outflows, we just have not clearly identified them yet. While some studies in the past revealed accelerated motions toward central cores (e.g., OMC-1, \citealt{hacar2017}), the more oscillatory and chaotic velocity structure of the innermost region discussed here does not allow us to infer any potential acceleration.


\section{Conclusions and summary}

Combining a unique multi-scale dataset from single-dish parsec-scale cloud coverage (MALT90) via intermediate-scale 0.25\,pc interferometric mosaic observations (ALMA-IMF) to the smallest-scale $<1000$\,au high-resolution ALMA data (DIHCA plus configuration 10 data), we can follow the gas flow from the infrared dark filament around the G351 star-forming region down to the innermost cores. 

While the large-scale data are only indicative of a gas flow, the intermediate-scale ALMA-IMF data clearly reveal filamentary structures leading from the north and south toward the central core. The red- and blue-shifted velocities of the gas are consistent with those from the largest spatial scales, hence indicating that indeed the flow starts on several parsec  scales and continues toward the center. Assuming that the velocity gradients are due to infalling filamentary gas structures, we can infer gas flow rates around $10^{-3}$\,M$_{\odot}$\,yr$^{-1}$,  consistent also with literature infall and accretion flow studies (section \ref{intermediate-scale}). Such high infall rates are needed to form high-mass stars within a few 100.000\,yrs (e.g., \citealt{mckee2003,bonnell2007,hoare2007,tan2014,motte2018}). This infalling gas is feeding a central region of roughly 10$^4$\,au size that hosts at least a dozen mm cores. Hence, the infall rates measured here correspond to the combined infall rates feeding all central cores and protostars. These rates are consistent with infall rates measured toward one of the central cores by absorption lines on the order $10^{-4}$ to $10^{-3}$\,M$_{\odot}$\,yr$^{-1}$.

Going toward the central $\sim$10$^4$\,au of the region, our combined multi-configuration very-high-angular-resolution dataset ($0.027''\times 0.022''$ or 50\,au resolution) reveals filamentary structures connecting the many cores identified previously. While the blue- and red-shifted gas at the edges of this central area agree well with the velocity structures of the intermediate-scale filaments, going toward the very center of the region, the kinematics become less ordered exhibiting often several velocity components, even at individual positions. We assign this more complex velocity structure in the center to an interplay of the converging infalling gas streams and at the same time feedback processes like molecular outflows from the central cores.

In summary, this multi-scale kinematic study of the prototypical infrared dark cloud and hot core region G351 reveals and quantifies the hierarchical gas flow from the large-scale cloud down to the innermost cores. While we find a relatively ordered converging gas flow from parsec down to 0.05\,pc scale, the innermost region reveals more complex velocity structures, likely caused by the combination of the converging gas flow with feedback processes from the central protostars.
 
\begin{acknowledgements}
We like to thank the referee for the insightful comments improving the quality of the paper. PS was partially supported by a Grant-in-Aid for Scientific Research (KAKENHI Number JP22H01271 and JP23H01221) of JSPS. PS was supported by Yoshinori Ohsumi Fund (Yoshinori Ohsumi Award for Fundamental Research). FO acknowledges the support of the NAOJ ALMA Joint Scientific Research Program grant No. 2024-27B, and from the National Science and Technology Council (NSTC) of Taiwan grants NSTC 112-2112-M-007-041 and NSTC 112-2811-M-007-048. AG acknowledges support from the NSF under grants AAG 2008101 and CAREER 2142300.

\end{acknowledgements}

\end{document}